\documentclass[aps,prd,groupedaddress,nofootinbib,floatfix]{revtex4}

\usepackage{graphicx,amsmath}
\usepackage{yfonts}
\DeclareGraphicsExtensions{.eps, .eps, .jpg, .png}

\begin{document}




\title{Schwinger effect in compact space: a real time calculation}



\author{Yue Qiu}

\email[]{yqiu.edu}
\author{Lorenzo Sorbo}

\email[]{sorbo@physics.umass.edu}

\affiliation{Amherst Center for Fundamental Interactions, Department of Physics, University of Massachusetts, Amherst, MA 01003, U.S.A.}


\date{\today}

\begin{abstract}
We compute the discharging rate of a uniform electric field due to Schwinger pair production in $(1+1)$-dimensional scalar electrodynamics with a compact dimension of radius $R$. Our calculation is performed in real time, using the in-in formalism. For large compactification radii, $R\to\infty$, we recover the standard non compact space result. However, other ranges of values of $R$ and of the  mass $m$  of the charged scalar give rise to a richer set of behaviors. For $R\gtrsim{\cal O}(1/m)$ with $m$ large enough, the electric field oscillates in time, whereas for $R\to 0$ it decreases in steps. We discuss the origin of these results. 
\end{abstract}

\pacs{98.80.Cq, 98.80.Qc}

\maketitle


\section{Introduction}%

The possibility of creating matter in the presence of a strong external field is a remarkable feature of relativistic quantum fields. The earliest and possibly most studied example of such phenomenon is the Schwinger effect~\cite{Sauter:1931zz,Schwinger:1951nm}. In this process, an external electric field accelerates charged virtual particles, extracting them from the vacuum and turning them into real pairs. The pair of produced particles generates an electric field that opposes the field responsible for their creation, reducing its intensity. The energy in the electric field  lost this way is thus converted into the energy in the particles.

The rate of pair production can be computed using different techniques, such as the early tunneling calculation of Sauter~\cite{Sauter:1931zz}, Schwinger's proper time method~\cite{Schwinger:1951nm}, instantons~\cite{Brown:1987dd} and real time techniques~\cite{Nikishov:1969tt,Kluger:1998bm,Tanji:2008ku,Garriga:2012qp}. 
Schwinger effect in compact spaces has received significant attention only in more recent years. In reference~\cite{Brown:2015kgj} the system was studied using the instanton formalism, and it was argued that for small compactification radii the expression of the rate of pair production changes significantly from the one found in the non compact case. The work~\cite{Draper:2018lyw} discussed the effect from the point of view of the effective lower-dimensional theory. The paper~\cite{Nagele:2018egu} studied numerically the unwinding of the electric flux as the produced pair circles multiple times the compact space.  Schwinger effect at finite temperature has received more attention~\cite{Medina:2015qzc,Brown:2015kgj,Gould:2017fve,Gould:2018ovk,Korwar:2018euc,Draper:2018lyw}, and is related to the spatially compact case as finite temperature effects can be captured by compactifying euclidean time.

In the present paper we study the Schwinger effect in compact space by directly computing, for the first time, the number of produced pairs as a function of time. More specifically, we compute the change, to leading order, in the intensity of the electric field induced by the current of produced pairs. We perform our calculation using the in-in formalism~\cite{Maldacena:2002vr,Weinberg:2005vy}, that at leading order corresponds to the calculation of Bogolyubov coefficients. As a consequence, our calculation parallels the one performed, for the non compact case,  in~\cite{Nikishov:1969tt,Kluger:1998bm,Tanji:2008ku,Garriga:2012qp}.  The system considered throughout the paper is scalar electrodynamics in $1+1$ spacetime dimensions, where the scalars have mass $m$ and charge $e$, and we denote by $E$ the magnitude of the background field.

Compactification of the spatial dimension implies that we can write the correction to the electric field as a series of contributions from the Kaluza-Klein modes of the charged field. The series is divergent, a consequence of the assumption that the background field is constant, so it had an infinite time to produce pairs. We isolate the divergent part, leaving the physical effect in the form of a convergent series, that can be in general evaluated numerically. Besides the renormalization associated to the infinite duration of the process, evaluation of the rate of pair production requires a different, more subtle kind of subtraction. Already in the non compact case, a direct computation of the rate of pair production in the in-in formalism gives a nonvanishing result even in the decoupling limit where the mass $m$ of the produced particles diverges. This unphysical behavior is taken care of by using the formalism of the Bogolyubov coefficients, where two-point functions are computed in terms of the normal-ordered ladder operators defined in the far future. Such an operation is equivalent to subtracting, from the two-point functions computed in terms of the exact mode functions, the same two-point functions written in terms of the mode functions evaluated in the adiabatic approximation~\cite{Birrell:1982ix,Parker:2009uva}.

While in the non compact case the electric field discharges at a uniform rate, the evolution of the electric field has a less trivial behavior in the small radius regime $R\to 0$. As one can see in Figure~\ref{fig:smallr}, the field is discharged by the effect of pair production in a stepwise fashion. While this is inconsistent with the assumption of a time-translation invariant background that would call for a uniform rate, we will argue in Section~\ref{subsec:VA} that this behavior is not unusual for a process of particle creation.

Another interesting behavior is that found in the regime $eER^2= {\cal O}(1)$, $mR\gtrsim {\cal O}(1)$.  For this choice of parameters, the electric field performs oscillations that, even if their amplitude is exponentially small, are parametrically larger than its decrease due to pair production -- see Figure~\ref{fig:mr12}.  As we discuss in Section~\ref{subsec:VB}, this behavior confirms the analysis of~\cite{Draper:2018lyw}. In particular, the oscillations are a direct consequence of the shape of the effective potential  of the electric field obtained in the dimensionally reduced theory, after integrating out the modes of the charged field. In other words, in this regime the effect on the evolution of the electric field is dominated by the effects of virtual pairs of scalar quanta, and not by the creation of actual pairs of particles. Our results thus confirm the interpretation~\cite{Draper:2018lyw} of the statement of~\cite{Brown:2015kgj} that, for small radii, the electric field evolves at a rate whose expression is different from the one found in the non compact theory. More specifically, while the expression of the {\em net} rate of pair production is unchanged, the amplitude of the {\em oscillations} of the electric field  is larger, and  matches the  amplitude of the effect discussed in~\cite{Brown:2015kgj}.

The plan of the paper is as follows. In Section~\ref{sec:noncompact} we set up our formalism and re-obtain, using the in-in formalism,  well known results on the rate of Schwinger pair production in non compact space. In Section~\ref{sec:compact} we present new formulae for the rate of pair production in real time in compact space. Details of the calculation are presented in two Appendices. In Section~\ref{sec:comp_results} we analyze,  in some representative regimes, the behavior of the formulae found in the previous Section. We discuss our results in Section~\ref{sec:discussion}.

\section{Non compact case}%
\label{sec:noncompact}

Before investigating case of a compact spatial dimension, here we review  the real-time analysis of the Schwinger effect in $1+1$ non compact dimensions. Our Lagrangian for scalar electrodynamics reads
\begin{eqnarray}
{\cal L}&=&-\frac{1}{4}F_{\mu\nu}F^{\mu\nu}-\big(\partial_\mu-ieA_\mu\big)\phi\,\big(\partial^\mu+ieA^\mu\big)\phi^*-m^2\,|\phi|^2\nonumber\\
&=&\frac{1}{2}\dot{A}^2+|\dot\phi|^2-|\phi'|^2-m^2\,|\phi|^2+ie\,A\,(\phi\,\phi'{}^*-\phi'\,\phi^*)-e^2\,A^2\,|\phi|^2\,,\label{Lagrangian}
\end{eqnarray}
where we have chosen the gauge $A_{\mu}=(0, A)$~\cite{Nagele:2018egu} and where a overdot and a prime denote, respectively, a time and a space derivative. We then decompose the gauge field $A$ into a background yielding a constant electric field $E$ and a perturbation
\begin{eqnarray}\label{gaugeA}
A(x,\,t)=-Et+\delta A(x,\,t)\,,
\end{eqnarray}
so that the Lagrangian~(\ref{Lagrangian}) takes the form
\begin{eqnarray}\label{LagrFI}
{\cal L}\equiv{\cal L}_F+{\cal L}_I&=&\frac{1}{2}\delta\dot{ A}^2+|\dot\phi|^2-|\phi'|^2-m^2\,|\phi|^2-ieEt\,(\phi\,\phi'{}^*-\phi'\,\phi^*)-e^2\,E^2t^2\,|\phi|^2\nonumber\\
&&+ie\,\delta A\,(\phi\,\phi'{}^*-\phi'\,\phi^*)-e^2\,\delta A^2\,|\phi|^2+2e^2\,\delta A\,Et\,|\phi|^2\,,
\end{eqnarray}
where ${\cal L}_F$ is the free part of the Lagrangian, given in the first line of eq.~(\ref{LagrFI}), whereas ${\cal L}_I$ denotes the interaction terms, given in the second line.

\subsection{The correction to the electric field in the in-in formalism} 

We now quantize the fields $\phi$ and $\delta A$ and compute the correction to the electric field to leading order in the in-in formalism. As we will show in Section~\ref{subsec:connect} below, the resulting correction to the electric field accounts for the creation of pairs of charged scalars through the Schwinger mechanism. 

The free equation for $\delta A$ is just $\delta\ddot{A}=0$, that has the general solution $\delta A(x,\,t)=\hat{c}(x)+\hat{d}(x)\,t$, where $\hat{c}(x)$ and $\hat{d}(x)$ are operators. Canonical quantization requires
\begin{align}
[\hat{c}(x)+\hat{d}(x)\,t,\,\hat{d}(y)]=i\delta(x-y)\,,\qquad [\hat{c}(x)+\hat{d}(x)\,t,\,\hat{c}(y)+\hat{d}(y)\,t]=[\hat{d}(x),\,\hat{d}(y)]=0\,,
\end{align}
that is satisfied by imposing the commutation relations
\begin{align}\label{eq:cd_commut}
[\hat{c}(x),\,\hat{d}(y)]=i\delta(x-y)\,,\qquad [\hat{c}(x),\,\hat{c}(y)]=[\hat{d}(x),\,\hat{d}(y)]=0\,.
\end{align}

Next, we decompose the field $\phi$  as
\begin{align}
\phi(x,\,t)\equiv \int\frac{dk}{\sqrt{2\pi}}\,e^{ikx}\,\hat{\phi}(k,\,t)=\int\frac{dk}{\sqrt{2\pi}}\,e^{ikx}\,\Big(\phi(k,\,t)\,\hat{a}_k+\phi(-k,\,t)^*\,\hat{b}^\dagger_{-k}\Big)\,,
\end{align}
where the mode function $\phi(k,t)$ solves the  equation of motion, derived from the free Lagrangian ${\cal L}_F$,
\begin{equation}
\ddot\phi(k,\,t)+\Big[(k+eEt)^2+m^2\Big]\phi(k,\,t)=0\,.
\end{equation}
The general solution of this equation can be written in terms of  parabolic cylinder functions
\begin{align}
\phi(k,\,t)=C_1\,D_{i\frac{m^2}{2\,eE}-\frac{1}{2}}\left((k+eEt)\sqrt{\frac{2}{eE}}e^{-i\pi/4}\right)+C_2\,D_{i\frac{m^2}{2\,eE}-\frac{1}{2}}\left(-(k+eEt)\sqrt{\frac{2}{eE}}e^{-i\pi/4}\right)\,,
\end{align}
where the integration constants $C_{1,\,2}$ can be fixed, after imposing the commutation relations $[\hat{a}_k,\,\hat{a}_{k'}^\dagger]=[\hat{b}_k,\,\hat{b}_{k'}^\dagger]=\delta(k-k')$, by requiring positive frequency modes at early times, when the WKB approximation can be applied
\begin{eqnarray}\label{def:phiad}
\phi(k,\,t\rightarrow-\infty)
\to \phi^{\rm ad}(k,t)
\equiv
\frac{1}{\sqrt{2\omega}}\,e^{-i\int \omega  dt }\approx \frac{1}{\sqrt{2|eEt|}}\,e^{i\frac{1}{2eE}(k+eEt)^2}\,|k+eEt|^{i\frac{m^2}{2\,eE}}\,.
\end{eqnarray}
This gives
\begin{align}
\phi(k,\,t)=\frac{e^{-\frac{\pi m^2}{8eE}}}{(2\,eE)^{1/4}}\,D_{i\frac{ m^2}{2eE}-\frac{1}{2}}\left(-(k+eEt)\sqrt{\frac{2}{eE}}e^{-i\pi/4}\right)\,.
\label{ncphi}
\end{align}

We compute the correction to electric field using the in-in formalism~\cite{Maldacena:2002vr,Weinberg:2005vy}, in which the expectation value of an operator ${\cal{O}}(t)$ is given by
\begin{eqnarray}\label{inin_expansion}
\langle {\cal O}(t)\rangle=\sum_N(-i)^N\int^t dt_1 ...\int^{t_{N-1}} dt_N\langle [[...[{\cal O}_{\rm free}(t),\,H_{\rm int}(t_1)],...],H_{\rm int}(t_N)]\rangle\,,
\end{eqnarray}
where $H_{\rm int}(t)$ denotes the interaction Hamiltonian, and ${\cal O}_{\rm free}$ represents the operator ${\cal O}$ computed in terms of the free mode functions.

In our case, the operator we are interested in is the correction to the electric field, $\delta E(x,t)\equiv -\delta\dot{A}(x,t)$. To lowest order in the in-in expansion, we only need to consider the cubic part of the interaction Hamiltonian,
\begin{eqnarray}
H^{(3)}_{\rm int}=-e\int dy\int\frac{dp\,dq}{(2\pi)}e^{i(p-q)y}\big[(p+q)+2\,eEt\big]\,\delta A(y)\,\hat{\phi}(p)\,\hat{\phi}^{\dagger}(q)\,.
\end{eqnarray}
Hence the first order correction to the electric field reads
\begin{eqnarray}
\langle \delta E(x,\,t)\rangle_{(1)}
&=&-ie\int^t dt_1\int dy\int\frac{dp\,dq}{(2\pi)}e^{i(p-q)y}\langle \Big[\delta \dot{A}(x,\,t),\,\delta A(y,\,t_1)\,\hat{\phi}(p,t_1)\,\hat{\phi}^{\dagger}(q,t_1)\Big]\rangle\big((p+q)+2\,eEt_1\big)\nonumber\\
&=&-e\int^t dt_1\int\frac{dp\,dq}{(2\pi)}e^{i(p-q)x} 
\langle\hat{\phi}(p,t_1)\,\hat{\phi}^{\dagger}(q,t_1)\rangle
\big[(p+q)+2\,eEt_1\big]\,,
\label{main}
\end{eqnarray}
that could have also been obtained by integrating Gauss' law.

\subsection{The two-point function of the scalar. Vacuum subtraction}
\label{sec:vacsub}

We must now compute the two point function $\langle\hat{\phi}(p,t)\,\hat{\phi}^{\dagger}(q,t)\rangle$. Since this quantity does not vanish even when the system is in its vacuum, we need to subtract its vacuum component.  We perform this subtraction by using the standard method of the Bogolyubov coefficients (see~\cite{Kluger:1998bm} for a nice discussion in this context). First, we define the Bogolyubov coefficients $\alpha(k,t)$ and $\beta(k,t)$ in such a way that

\begin{eqnarray}\label{eq:deco_phiphiad}
\phi(k,\,t)=\alpha(k,t)\, \phi^{\rm ad}(k,t)+\beta(-k,t)\,\phi^{\rm ad}(-k,t)^*\,,
\end{eqnarray}
where $\phi^{\rm ad}(k,t)$ is defined in eq.~(\ref{def:phiad}). It is worth stressing here that we are assuming that the function $\phi^{\rm ad}(k,t)$ is evaluated to leading order in the adiabatic approximation, but that in general higher order expressions might be more appropriate~\cite{Berry,Ritus,Dumlu:2010ua,Dabrowski:2014ica,Dabrowski:2016tsx} . Eq.~(\ref{eq:deco_phiphiad}) implies that $\alpha(k,t)$ and $\beta(k,t)$ are constant in the adiabatic regime. We then define a new set of operators $\hat{\textswab{a}}_{k}(t)$, $\hat{\textswab{b}}_{k}(t)$ so that
\begin{eqnarray}
\hat{\phi}(k,\,t)=\phi^{\rm ad}(k,t)\, \hat{\textswab{a}}_{k}(t)+\phi^{\rm ad}(-k,t)^*\, \hat{\textswab{b}}_{-k}(t)^{\dagger}\,.
\end{eqnarray}
This is equivalent to redefining
\begin{align}
 \hat{\textswab{a}}_{k}(t)&=\alpha(k,t) \,\hat{a}_k+\beta(k,t)^*\,\hat{b}^\dagger_{-k}\nonumber\\
 \hat{\textswab{b}}_{k}(t)&=\beta(k,t)^*\,\hat{a}_{-k}^{\dagger}+\alpha(k,t)\,\hat{b}_{k}\,.
\end{align}

We now impose that the quantity $\langle\hat{\phi}(p,t_1)\,\hat{\phi}^{\dagger}(q,t_1)\rangle$ is computed as the expectation value on the initial state vacuum (annihilated by $\hat{a}_{k}$ and $\hat{b}_{k}$) after normal ordering the $\hat{\textswab{a}}_{k}(t)$ and $\hat{\textswab{b}}_{k}(t)$ operators. This prescription generalizes the one for the occupation number $N_k=\langle \hat{\textswab{b}}_{k}^\dagger\,\hat{\textswab{b}}_{k}\rangle=|\beta(k,t)|^2$ to a generic bilinear in the field.

Using this prescription we obtain
\begin{align}
&\langle\hat{\phi}(p,t)\,\hat{\phi}^{\dagger}(q,t)\rangle=\delta(p-q)\left[|\phi(q,t)|^2-|\phi^{\rm ad}(q,t)|^2\right]\,.
\end{align}

This method therefore leads to results that are equivalent to those obtained through adiabatic regularization~\cite{Birrell:1982ix,Parker:2009uva}. Using the expression~(\ref{def:phiad}) of $\phi^{\rm ad}(k,t)$ obtained to leading order in the adiabatic approximation, the renormalized result is 
\begin{eqnarray}
 \langle\hat{\phi}(p,t_1)\hat{\phi}^{\dagger}(q,t_1)\rangle
 =\delta(p-q)\,\left[|\phi(p,t_1)|^2-\frac{1}{2\sqrt{(p+eEt_1)^2+m^2}}\right]\,,\label{twopoint}
\end{eqnarray}
that, substituted into~(\ref{main}), gives the final expression
\begin{eqnarray}\label{main2}
\langle \delta E(x,\,t)\rangle_{(1)}&=&-2e\int^t dt_1\int\frac{dp}{2\pi} \big|\phi(p,t_1)\big|^2\big(p+eEt_1\big)+e\int^t dt_1\int\frac{dp}{2\pi} \frac{p+eEt_1}{\sqrt{(p+eEt_1)^2+m^2}}\,. 
\end{eqnarray}

\subsection{Result}

The integral in eq.~(\ref{main2}) can be computed exactly, but we do not need to perform this calculation. It is  easier to extract the physically relevant result by observing that $p$ and $t$ appear in the integrands in eq.~(\ref{main2}) always in the combination $(p+eEt)$, so that derivatives with respect to $t$ can be easily traded for derivatives with respect to $p$: $\partial/\partial t=eE\,\partial/\partial p$. The second derivative of eq.~(\ref{main2}) then reads
\begin{eqnarray}\label{quasi_final_noncompact}
\langle \delta\ddot{E}(x,\,t)\rangle_{(1)}&=&
\lim_{\Lambda_\pm\rightarrow+\infty}
\left[-\frac{e}{\pi} \int_{-\Lambda_-}^{\Lambda_+} dp\frac{d}{dt}\left\{\big|\phi(p+eEt)\big|^2\big(p+eEt\big)\right\}
+\frac{e}{2\pi} \int_{-\Lambda_-}^{\Lambda_+} dp\frac{d}{dt}\left\{\frac{p+eEt}{\sqrt{(p+eEt)^2+m^2}} \right\}\right]\nonumber\\
&=&
-\frac{e^2E}{\pi} \frac{e^{-\frac{\pi m^2}{4eE}}}{(2\,eE)^{1/2}}\,
\lim_{\Lambda_\pm\rightarrow+\infty}
\Big[\big|D_{i\frac{\pi m^2}{2eE}-\frac{1}{2}}((p+eEt)\sqrt{\frac{2}{eE}}e^{3i\pi/4})\big|^2\big(p+eEt\big)\Big]_{p=-\Lambda_-}^{p=\Lambda_+}
+\frac{e^2E}{\pi}\,, 
\end{eqnarray}
where we have regularized the $dp$ integrals by setting the integration range on $(-\Lambda_-,\Lambda_+)$, and where we have used the fundamental theorem of integral calculus in the second line.

Using the asymptotic behavior of the parabolic cylinder function, see e.g. eq.~9.246 of~\cite{gradshteyn}, we thus obtain
\begin{eqnarray}\label{final_noncompact}
\langle \delta \ddot{E}(x,\,t)\rangle_{(1)}=-\frac{e^2E}{\pi}\left[\left(e^{-\frac{\pi m^2}{eE}}+\frac{1}{2}\right)-\left(-\frac{1}{2}\right)\right]+\frac{e^2E}{\pi}=-\frac{e^2E}{\pi}\,e^{-\pi \frac{m^2}{eE}}\,.
\end{eqnarray}

Two comments are in order. First, if we did not subtract the adiabatic part of $ \langle\hat{\phi}(p,t_1)\hat{\phi}^{\dagger}(q,t_1)\rangle$ (which results in the last term in the second line of eq.~(\ref{quasi_final_noncompact})), the resulting rate~(\ref{final_noncompact}) would not vanish in the decoupling limit $m\to\infty$. This shows the need for the subtraction of the adiabatic part of the two point function of $\phi$.

Second, the equation above gives the second time derivative of $\langle {\delta E}(x,\,t)\rangle_{(1)}$. This means that $\langle \delta {E}(x,\,t)\rangle_{(1)}$ contains two integration constants, that are however unrelated to rate of pair production and thus are uninteresting for us. The first of these constants is related to the initial value of the background electric field, the second one is related to the initial value of the number of charged particles, that are subsequently accelerated and decrease the background field at a rate that is linear in time. Neither of these integration constants is related to the rate of pair production, that is fully captured by eq.~(\ref{final_noncompact}), as we now discuss.

\subsection{Connecting $\ddot{E}$ to to the rate of pair production}%
\label{subsec:connect}

We will now connect the rate~(\ref{final_noncompact}) of change of the electric field to the rate of  production of pairs of $\phi$ particles.

If a particle of mass $m$ and charge $e$ is subject to a uniform field $E$, then its velocity is given by $v(t)=\frac{eE(t-t_0)}{\sqrt{m^2+(eE(t-t_0))^2}}$, where $t_0$ is the time at which the particle is at rest.

Let now $dn_\pm(t_0)$ be the number density of particles with charge $\pm e$ created at rest between the times $t_0$ and $t_0+dt_0$. Then the element of current associated to those particles and evaluated at time $t$ reads
\begin{align}
dJ(t,\,t_0)=e\,dn_+(t_0)\frac{eE(t-t_0)}{\sqrt{m^2+(eE(t-t_0))^2}}-e\,dn_-(t_0)\frac{(-e)E(t-t_0)}{\sqrt{m^2+(eE(t-t_0))^2}}\,.
\end{align}
Using the fact that the production rates are the same for both particles and antiparticles and that they are constant in time, so that $dn_\pm(t_0)=\dot{n}\,dt_0$ with $\dot{n}=$constant, we get
\begin{align}
dJ(t,\,t_0)=2\dot{n}\frac{e^2E(t-t_0)}{\sqrt{m^2+(eE(t-t_0))^2}}\,dt_0\,,
\end{align}
and finally the current at time $t$ is
\begin{align}
J(t)=\int_{t_{\rm{in}}}^{t}\frac{dJ(t,\,t_0)}{dt_0}\,dt_0=2\frac{\dot{n}}{E}\left(\sqrt{m^2+e^2E^2(t-t_{\rm{in}})^2}-m\right)\to 2\,\dot{n}\,e\,(t-t_{\rm {in}})\,,
\end{align}
where we assumed that the process started long ago,  $eE(t-t_{\rm {in}})\gg m$.

Then, from the Maxwell equations $\partial_\mu F^{\mu\nu}=J^\nu$ we obtain $\dot{E}=-J\Longrightarrow \ddot{E}=-\dot{J}=-2\,\dot{n}\,e$. This finally gives
\begin{align}
\dot{n}=-\frac{\ddot{E}}{2e}=-\frac{\delta\ddot{E}}{2e}=\frac{eE}{2\pi}\,e^{-\pi \frac{m^2}{eE}}\,,
\end{align}
in agreement with the standard result.

\section{Compact case}%
\label{sec:compact}

Let us now consider a system where space is compactified, with the identification $x\approx x+2\pi R$. The procedure for the compact case is similar to  that presented above in the non compact regime.  As we will see, however, additional complications come from the fact we will not be able to use the fundamental theorem of integral calculus that allowed to simplify eq.~(\ref{quasi_final_noncompact}).

\subsection{Kaluza-Klein decomposition}

We decompose the gauge field as in the non compact case: $\delta A(x,t)=\hat{c}(x)+\hat{d}(x)\, t$, with $\hat{c}(x)$ and $\hat{d}(x)$ periodic functions satisfying the commutation relations of eq.~(\ref{eq:cd_commut}). For what concern the field $\phi$, we decompose it into Kaluza-Klein modes
\begin{eqnarray}
\phi(t,x)\equiv\sum_{n=-\infty}^\infty \frac{1}{\sqrt{2\pi R}}\,e^{inx/R}\,\hat{\phi}_n(t) 
=\sum_{n=-\infty}^\infty \frac{1}{\sqrt{2\pi R}}\,e^{inx/R}\,
\big[\phi_n(t)\,\hat{a}_n+\phi^*_{-n}(t)\,\hat{b}^\dagger_{-n}\big]\,,
\end{eqnarray}
where, analogously to eq.~(\ref{ncphi}), $\phi_n(t)$ is given by
\begin{eqnarray}
\phi_n(t)=\frac{e^{-\frac{\pi m^2}{8eE}}}{(2\,eE)^{1/4}}\,D_{i\frac{m^2}{2eE}-\frac{1}{2}}\Big(-(\tfrac{n}{R}+eEt)\sqrt{\tfrac{2}{eE}}e^{-i\pi/4}\Big)\,.
\end{eqnarray}
Then, substituting into eq.~(\ref{main}), one obtains
\begin{eqnarray}\label{eq:implicit_sum}
\langle \delta E(x,\,t)\rangle_{(1)}&=&-\frac{e}{\pi R}\int^t dt'\sum_n \big|\phi_n(t')
\big|^2\big(\frac{n}{R}+eEt'\big)
+\frac{e}{2\pi R}\int^t dt'\sum_n \frac{\frac{n}{R}+eEt'}{\sqrt{(\frac{n}{R}+eEt')^2+m^2}}\,.
\end{eqnarray}

\subsection{Result}

Since the transition to the compact case converts the integral in $dp$ in eq.~(\ref{main2}) into a series, we cannot use the trick -- based on the fundamental theorem of calculus -- used in eq.~(\ref{quasi_final_noncompact}) to compute $\langle\delta\ddot{E}\rangle$. Instead, we have to compute the series directly. An additional complication is that the series in eq.~(\ref{eq:implicit_sum}) are divergent. So we will use a different strategy. First, we take the time derivative $\langle \delta\dot{E}(x,\,t)\rangle_{(1)}$, that eliminates the time integral from eq.~(\ref{eq:implicit_sum}). Then we cut off the summation at some large $N_\pm>0$, $-N_-<n<N_+$. We use a Mellin-Barnes representation (see eq. 9.242.3 of~\cite{gradshteyn}) of the parabolic cylinder function. This allows to express the $n$-dependence of the first term on the right hand side of eq.~(\ref{eq:implicit_sum}) in the simple form of Hurwitz $\zeta$-like series, and the divergent part as $N_\pm\to\infty$ can be isolated. The remaining part of series can be resummed to a finite result and reverse-engineered using again the Mellin-Barnes representation of the parabolic cylinder functions. The details are presented in Appendix~\ref{app:maincalc}. 

As for the vacuum part, given by the second summation on the right hand side of eq.~(\ref{eq:implicit_sum}), the analysis  is simpler. Here we just state that, also in this case, we have to cut the sum off at $-N_-<n<N_+$. Then we can rewrite (Appendix~\ref{app:sigma} gives the details)
\begin{align}
\sum_{n=-N_-}^{N_+} \frac{\frac{n}{R}+eEt}{\sqrt{(\frac{n}{R}+eEt)^2+m^2}}=2\,eERt+4\,mR{}\sum_{n=1}^\infty \sin(2\pi n eERt)\,K_1(2\pi n mR)+{\rm {constant}}\,,
\end{align}
where $K_1$ is the modified Bessel function of second kind, and where the constant part is generally divergent as $N_\pm\to+\infty$, but does not depend on $t$.

Our final expression for the time derivative of the electric field, at first order in the perturbative expansion, reads
\begin{align}\label{eq:final_sum}
&\langle \delta \dot{E}(t)\rangle_{(1)}=-\frac{e}{2\pi R}\Bigg\{(1+2\,e^{-\pi m^2})\sum_{n=0}^\infty \left[\frac{\sqrt{2}(n+Rt-[Rt])}{R}\,e^{-\pi m^2/4}\left|D_{-1/2+im^2/2}\left(\frac{\sqrt{2}(n+Rt-[Rt])}{R}\,e^{-i\pi/4}\right)\right|^2-1\right]\nonumber\\
&-\sum_{n=0}^\infty \left[\frac{\sqrt{2}(n+1-Rt+[Rt])}{R}\,e^{-\pi m^2/4}\left|D_{-1/2+im^2/2}\left(\frac{\sqrt{2}(n+1-Rt+[Rt])}{R}\,e^{-i\pi/4}\right)\right|^2-1\right]\nonumber\\
&+2\left([Rt]-Rt\right)-4\,mR{}\sum_{n=1}^\infty \sin(2\pi n Rt)\,K_1(2\pi n mR)\Bigg\}-\frac{e}{\pi R}\,e^{-\pi m^2}\,[Rt]+{\rm constant}\,,
\end{align}
where, in order to keep the a lighter notation, we have set $eE=1$ (and we will do this in the remainder of this Section and in Section~\ref{sec:comp_results}),  where $[x]$ denotes the integer part of $x$, and where the constant term depends on the regulators $N_\pm$, and is generally divergent as $N_\pm\to\infty$.

Eq.~(\ref{eq:final_sum}) is our main result. It gives the rate of change of electric field due to Schwinger pair production in compact spaces for arbitrary values of the parameters, at first order in the in-in expansion. In the next Section we will study the behavior of this function.  Here, let us note a couple of general properties. First, all the  parts that are divergent as $N_\pm\to+\infty$ are independent on $t$. As discussed above, these divergences correspond to a possible nonvanishing initial charge distribution in the system and are uninteresting for us.  All the parts that are not explicitly dependent on $N_\pm$ are finite (in particular, this means that the summations appearing in eq.~(\ref{eq:final_sum}) are convergent).  Second, the nontrivial part of eq.~(\ref{eq:final_sum}) depends on $Rt-[Rt]$, so that, as a function of $t$, is periodic with a period $1/R$. The non periodic component of~$\langle \delta\dot{E}(t)\rangle_{(1)}$ is given by the term outside the curly brackets in eq.~(\ref{eq:final_sum}), $-\tfrac{e}{\pi}\,e^{-\pi m^2}\,\tfrac{[Rt]}{R}$.  This means that the {\em net} rate of particle production, averaged on timescales that are long with respect to the timescale $R^{-1}$, coincides with the non compact result.

\section{Results for compact case}%
\label{sec:comp_results}

We are now in position to analyze the behavior of our main observable, $\langle \delta\dot E(x,\,t)\rangle_{(1)}$, in various regimes.

\subsection{The decompactified limit, $R\to\infty$ with fixed $m$}

Let us first make sure that our result~(\ref{eq:final_sum}) converges to the non compact one, eq.~(\ref{final_noncompact}), in the limit $R\to\infty$. In this limit we note that the sums in the first two lines of eq.~(\ref{eq:final_sum}) can be approximated by integrals, where we introduce an integration variable $p=n/R$, so that
\begin{align}\label{eq:sum_large_r}
&\langle \delta\dot{E}(t)\rangle_{(1)}=-\frac{e}{2\pi}\Bigg\{(1+2\,e^{-\pi m^2})\int_{0}^\infty dp\,\left[\sqrt{2}(p+\tau/R)\,e^{-\pi m^2/4}\left|D_{-1/2+im^2/2}\left(\sqrt{2}(p+\tau/R)\,e^{-i\pi/4}\right)\right|^2-1\right]\nonumber\\
&-\int_0^\infty dp \left[\sqrt{2}(p+(1-\tau)/R)\,e^{-\pi m^2/4}\left|D_{-1/2+im^2/2}\left(\sqrt{2}(p+(1-\tau)/R)\,e^{-i\pi/4}\right)\right|^2-1\right]\nonumber\\
&-2\frac{\tau}{R}-4\,m{}\sum_{n=1}^\infty \sin(2\pi n \,Rt)\,K_1(2\pi n mR)\Bigg\}-\frac{e}{\pi R}\,e^{-\pi m^2}\,[Rt]+{\rm constant}\,,
\end{align}
where we have defined the quantity
\begin{align}
\tau\equiv Rt-[Rt]\,,\qquad 0\le \tau<1\,.
\end{align}
Next we note that $\tau/R\to 0$ in this limit, which implies that the first two lines in eq.~(\ref{eq:sum_large_r}) converge to a time-independent and irrelevant constant. Moreover, for what concerns the term involving the Bessel function $K_1$, one can use the asymptotic behavior $K_1(x)\approx \sqrt{\frac{\pi}{2x}}e^{-x}$ to show that that term vanish exponentially fast as $mR\to \infty$. As a consequence, using $\tfrac{[Rt]}{R}\to t$ as $R\to \infty$, and reinstating, just for this result, the factors of $eE$, we re-obtain the non compact result
\begin{align}\label{eq:sum_large_r_fin}
&\langle \delta\dot{E}(t)\rangle_{(1)}\xrightarrow[R\to\infty]{}-\frac{e^2E}{\pi} e^{-\pi\frac{m^2}{eE}}\,t+{\rm {constant}}\,.
\end{align}

\begin{figure}
\centering
\includegraphics[scale=.4]{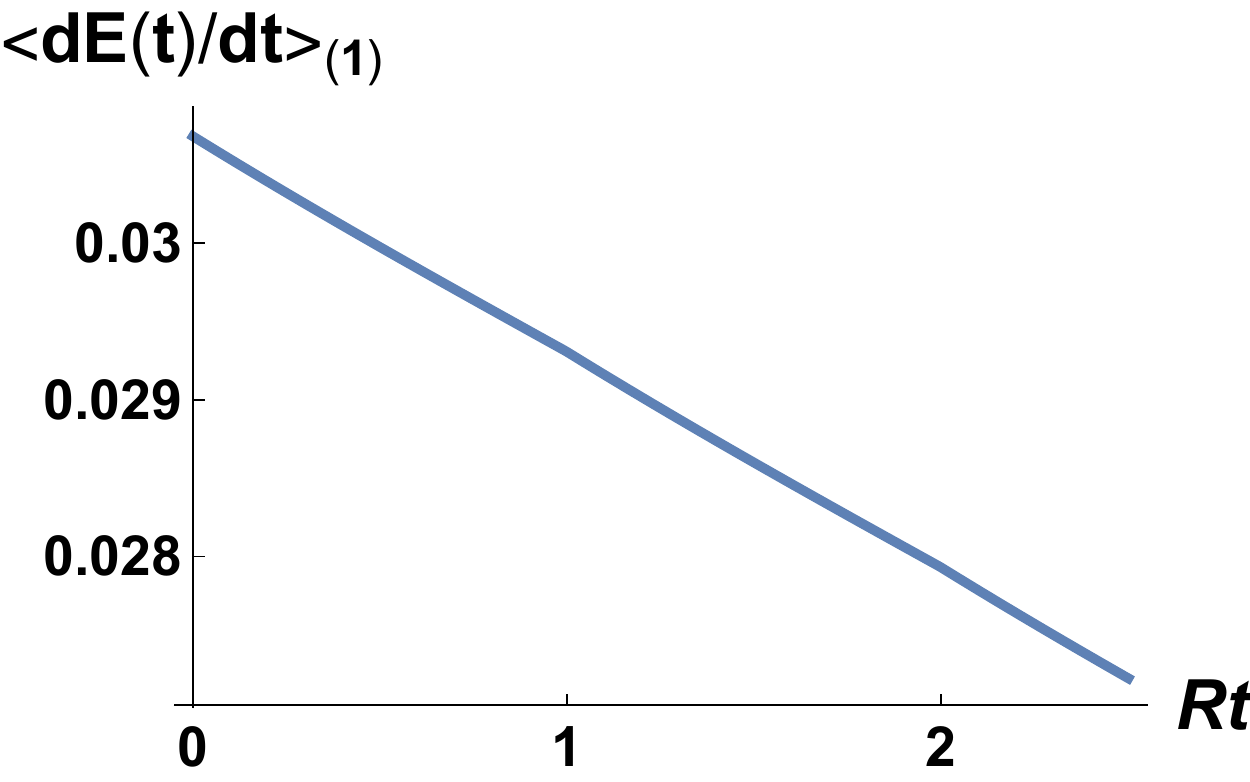} 
\caption{The quantity $\langle \dot{E}(t)\rangle_{(1)}$, in units of $e$ and for $eE=1$, as a function of $Rt$ evaluated numerically for $m=1$, $R=10$.  The result agrees with the analytical approximation~(\ref{eq:sum_large_r_fin}).}
\label{fig:larger}
\end{figure}

\subsection{The Kaluza-Klein decoupling limit, $R\to 0$ with fixed $m$}

For $R\to 0$, the Kaluza-Klein modes become infinitely heavy, so that we expect each Kaluza-Klein mode of the charged fields to be separately excited. To study this regime analytically we assume $\frac{\tau}{R}\gg 1$ and  $\frac{1-\tau}{R}\gg 1$, so that we are far enough from the instances of particle production that occur when $Rt$ crosses a integer value. We can then use the asymptotic expansion of the parabolic cylinder function for large arguments, which gives
\begin{align}
x\,e^{-\pi m^2/4}\left|D_{-1/2+im^2/2}\left(x\,e^{-i\pi/4}\right)\right|^2-1=-\frac{m^2}{x^2}+O(x^{-4})\,.
\end{align}
So we get, for $R\to 0$,
\begin{align}
&\langle \delta\dot{E}(t)\rangle_{(1)}\simeq \frac{m^2eR}{4\pi}\,\left(1+2\,e^{-\pi m^2}\right)\,\psi'(Rt-[Rt])-\frac{m^2eR}{4\pi}\psi'(1-Rt+[Rt])-\frac{e}{\pi R}\,e^{-\pi m^2}\,[Rt]+{\rm constant}\,,
\end{align}
where $\psi'$ denotes the derivative of the digamma function, $\psi(x)=\Gamma'(x)/\Gamma(x)$, $\psi'(x)=\sum_{n=0}^\infty(n+x)^{-2}$, and where we have used eq.~(\ref{eq:deltaSsmallR}). As $R\to 0$, the terms proportional to the $\psi'$ functions vanish, and we are left with the result
\begin{align}\label{eq:final_rto0}
&\langle \delta\dot{E}(t)\rangle_{(1)}\simeq -\frac{e}{\pi}\,e^{-\pi m^2}\frac{[Rt]}{R}\,,\qquad {\rm with}\qquad \frac{Rt-[Rt]}{R}\gg 1\,,\quad  \frac{1-Rt+[Rt]}{R}\gg 1\,,\qquad R\to 0\,.
\end{align}

\begin{figure}
\centering
\includegraphics[scale=.5]{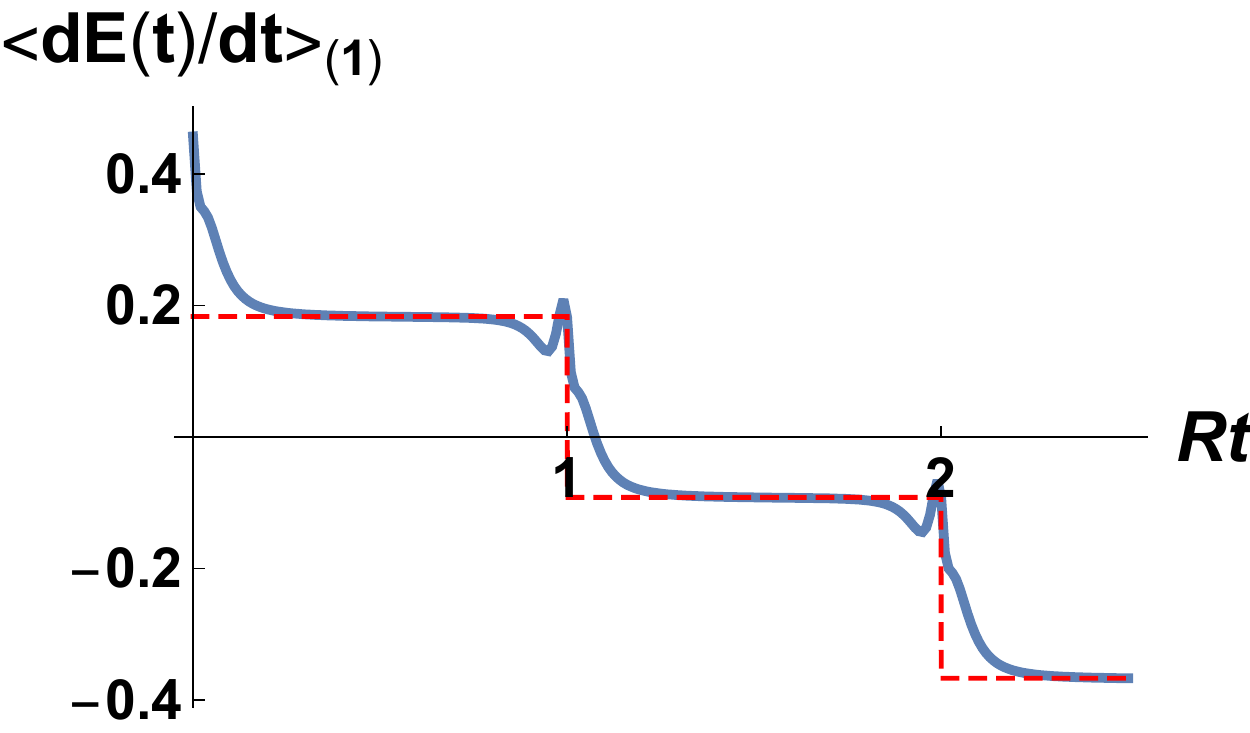} 
\hspace*{2cm}
\includegraphics[scale=.5]{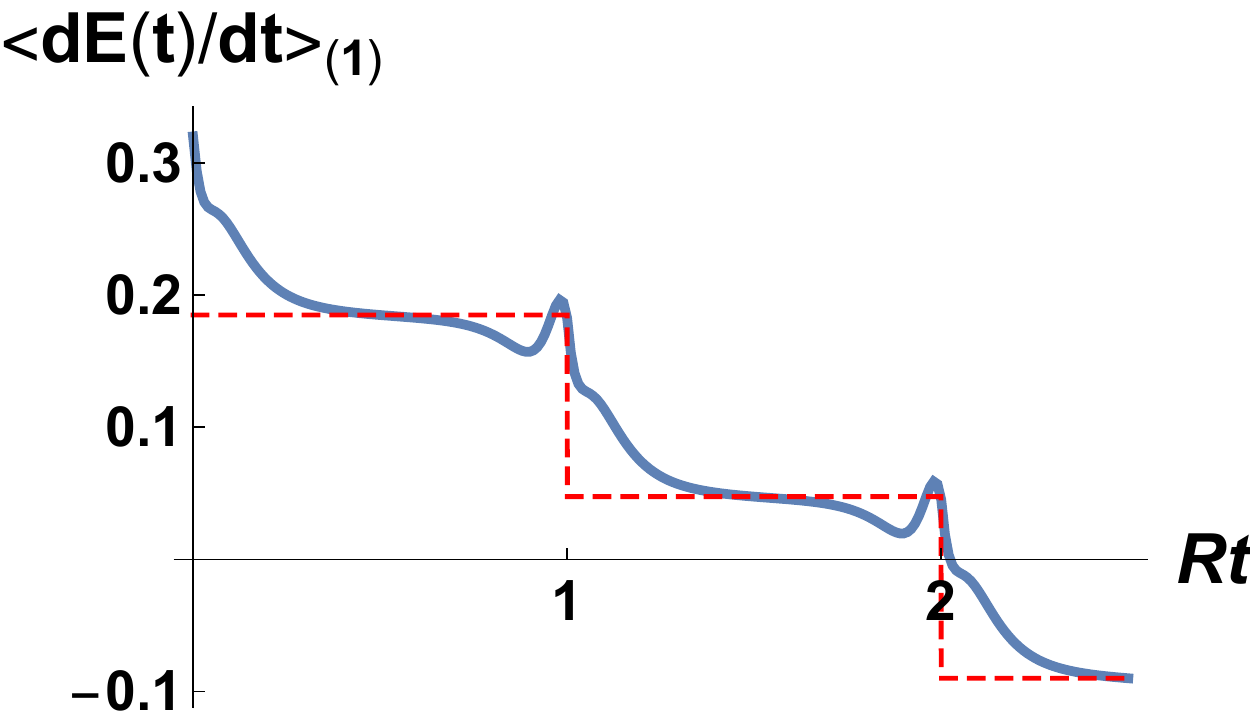} 
\caption{The quantity $\langle\dot{E}(t)\rangle_{(1)}$, in units of $e$ and for $eE=1$, as a function of $Rt$ evaluated numerically for $m=1$, $R=.05$ (left) and  $m=1$, $R=.1$ (right).  For $Rt$ far from integer values the result agrees with the analytical approximation~(\ref{eq:final_rto0}), shown in the red, dashed lines.}
\label{fig:smallr}
\end{figure}

As stated above, this result requires $Rt$ to be far from an integer value. To cover also the case where $Rt$ is close to integer we have to compute $\langle \delta\dot{E}(t)\rangle_{(1)}$ numerically. We show the results (in units of $e$) obtained for two choices of parameters in Figure~\ref{fig:smallr}. As one can see, eq.~(\ref{eq:final_rto0}) provides an excellent approximation of the exact result for $|\tfrac{Rt-[Rt]}{R}|\gg 1$, while when $Rt$ is close to an integer the amplitude of the correction to the electric field shows a non-trivial behavior.

\subsection{$R={\cal O}(1)$ with  $mR\gtrsim {\cal O}(1)$}
\label{subsec:rmgg1}

We conclude this Section analyzing the case in which $mR\gtrsim{\cal O}(1)$, which include the case $mR\gg 1$. We also assume $m\gtrsim 1$, which implies that the net rate of pair production, as given by the non periodic component of eq.~(\ref{eq:final_sum}) is negligible.  For this choice of parameters the distance in time $\sim 1/R$ between the events of production of different Kaluza-Klein modes of the matter field is comparable to or much smaller than the duration $\sim m$ of the individual events of particle production themselves (remember that we are setting $eE=1$ in this Section).  To see that particle production lasts a time $\sim m$, let us remember that this process occurs when the frequency of the mode functions is evolving non adiabatically. In our case, considering without loss of generality the zero mode of the field $\phi$, the frequency  reads $\omega=\sqrt{t^2+m^2}$. Nonadiabaticity is maximal when the quantity $|\dot\omega|/\omega^2$ is maximized, which turns out to be the case when $t=m/\sqrt{2}$.

\begin{figure}
\centering
\includegraphics[scale=.5]{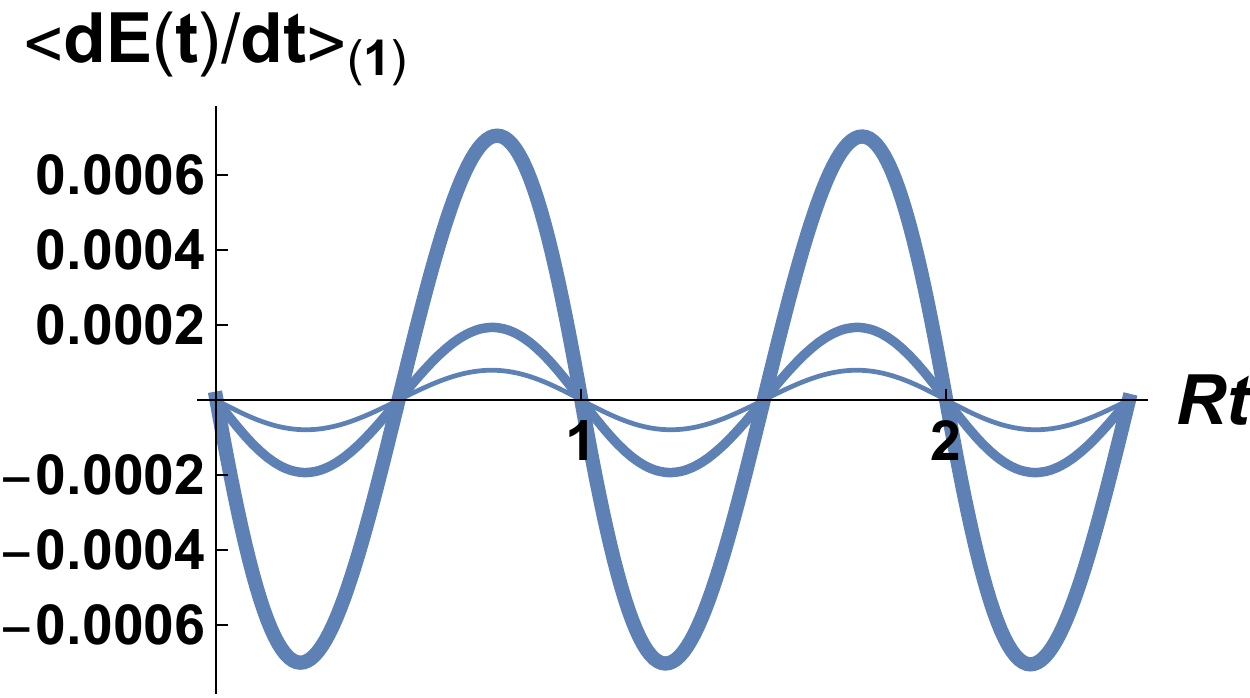} 
\hspace*{2cm}
\includegraphics[scale=.18]{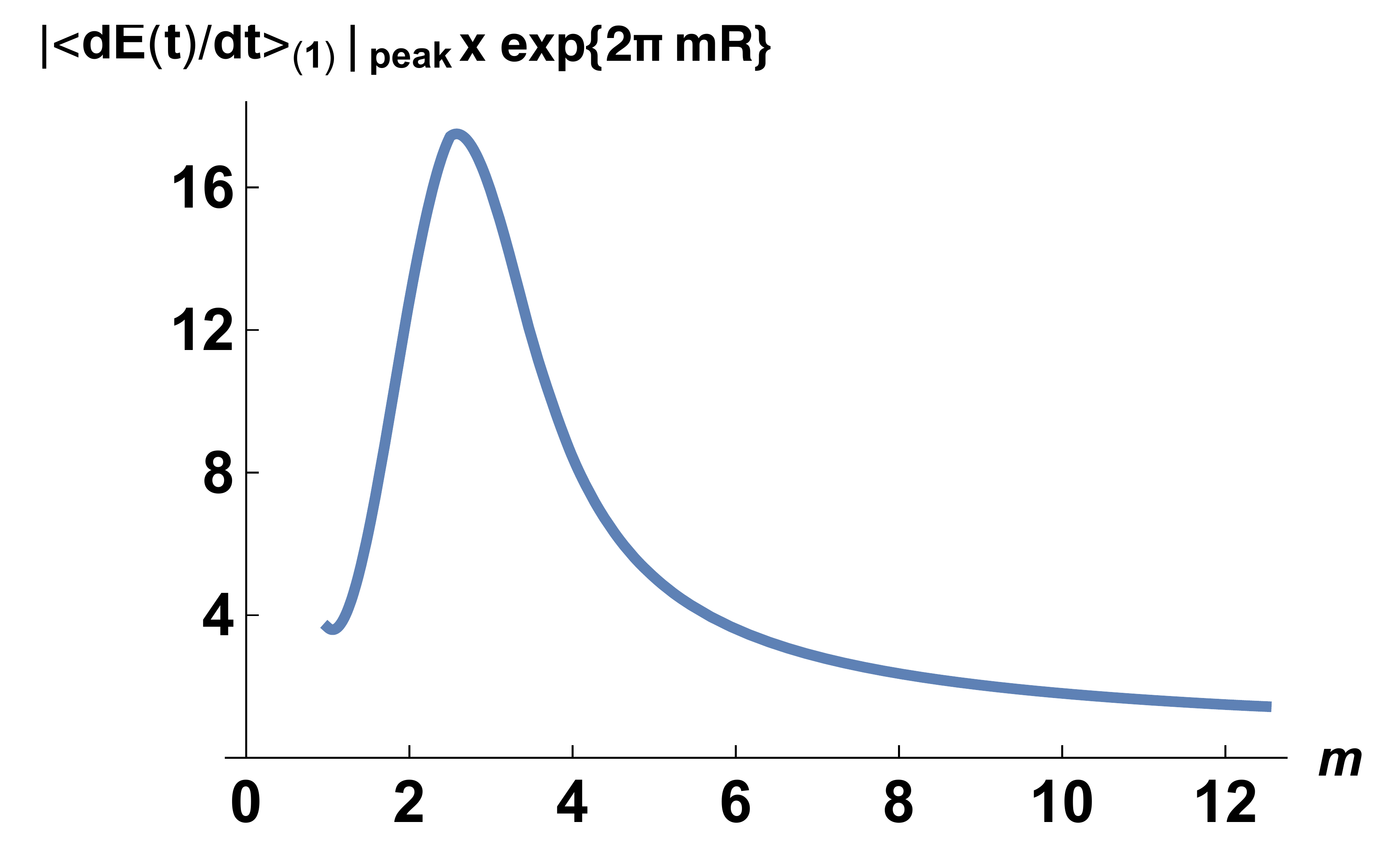}
\caption{Left panel: the quantity $\langle \dot{E}(t)\rangle_{(1)}$,  in units of $e$ and for $eE=1$,  as a function of $Rt$ evaluated numerically for $mR=1$.  The curves, from larger to smaller amplitude, correspond to $m=2,\,3,\,4$. Right panel: the amplitude of the oscillations of the quantity $\langle \dot{E}(t)\rangle_{(1)}$, multiplied by $e^{2\pi mR}$, as a function of $m$ and for $R=1$.}
\label{fig:mr12}
\end{figure}

Since for $R\gtrsim{\cal O}(1/m)$ there is always at least one Kaluza-Klein mode of the scalar field whose proper frequency is not evolving adiabatically, we find that the system is never in a fully adiabatic regime. Numerical analysis, indeed, shows a rather unusual behavior: the electric field performs sinusoidal oscillations. The amplitude of such oscillations, in the regime of large $mR\gg 1$, that can be seen to go as $e^{-2\pi mR}$. This behavior is apparent in the plots in Figure~\ref{fig:mr12}. For smaller values of $m$ and fixed $mR={\cal O}(1)$ (not shown) the behavior is similar, but the oscillations are superimposed to a decay due to the term proportional to $[Rt]$ at the end of eq.~(\ref{eq:final_sum}).

In Section~\ref{subsec:VB} below we discuss the origin of this behavior.

\section{Discussion and conclusions}
\label{sec:discussion}

In the previous Section we have analyzed the function $\langle \delta\dot{E}(t)\rangle_{(1)}$ in various regimes. We have seen that in the decompactification limit $R\to\infty$ we recover the standard result~(\ref{eq:sum_large_r}). In the opposite $R\to 0$ limit we have obtained a stepwise behavior, which seems inconsistent with the stationary nature of the background that would require a uniform rate of particle creation. Finally, in the case $R={\cal O}(1)$, $mR\gg 1$, the function $\langle \delta\dot{E}(t)\rangle_{(1)}$ also features an unexpected time dependence, showing oscillations whose amplitude is parametrically larger than the decrease of the electric field due to particle production. 

Let us discuss the latter two results separately.

\subsection{On the steplike behavior for $R\to 0$}
\label{subsec:VA}

The steplike behavior observed in Figure~\ref{fig:smallr} is surprising if one considers that the background system displays invariance under {\em continuous} time translations. Why is this continuous symmetry broken down to the discrete invariance under $t\mapsto t+1/(eER)$, and what does determine the exact time when particle production occurs? 

The fact that vacuum decay might break some of the background spacetime symmetries has been already discussed, for instance in~\cite{Garriga:2011we,Dine:2012tj,Garriga:2012qp}. Analogously to what was shown in~\cite{Garriga:2012qp}, which focused on the Lorentz invariance of the Schwinger phenomenon in non compact space, the {\em vacuum} of our system is invariant under time translation, as a time translation $t\mapsto t+\delta t$ is canceled by a gauge transformation $A\mapsto A-E\,\delta t$. Conversely, the addition of a constant to the gauge potential would determine a shift in the times of particle production. In compact space, a physical observable can have gauge dependence, as in the Aharonov-Bohm effect, which implies that gauge transformations can also be observable unless they meet certain conditions. In our case, gauge invariance is preserved if the particle picks up a phase equal to a multiple of $2\pi$, which corresponds to a time translation $\delta t$ is equal to a multiple of $\frac{1}{eER}$, and which in its turn corresponds to a shift of the Kaluza-Klein modes by an integer. Choosing a gauge, as we did, is equivalent to choosing an initial time for a particular observer who measures the particle productions, analogously to the choice of an observer's frame for bubble nucleation, as discussed in~\cite{Garriga:2012qp}.  Finally, it is worth noting that an exactly time-translational invariant electric field is an idealization -- if this state is decaying then it cannot have been around forever, otherwise it would have already completely decayed. An electric field must be turned on at some initial time, which will break the time translation symmetry manifestly. 

\subsection{On the oscillations for $eER^2={\cal O}(1)$, $m^2\gg eE$}
\label{subsec:VB}

The sinusoidal time dependence observed in the regime $eER^2={\cal O}(1)$, $m^2\gg eE$ can be explained by invoking the $(0+1)$-dimensional effective description of the model, as done in~\cite{Draper:2018lyw}. In fact, in this regime particle production is negligible and one can just study the theory that results by integrating out the heavy Kaluza-Klein modes of the bulk charged matter. Such an operation leads to a $(0+1)$-dimensional  Lagrangian that in the limit $mR\gg1$ reads
\begin{align}\label{eq:eff_0p1lagr}
L_{0+1,\,{\rm eff}}\simeq \pi R\dot{A_0}(t)^2+\frac{m^{1/2}}{\pi R^{1/2}}e^{-2\pi mR}\,\cos(2\pi R eA_0(t))\,,
\end{align}
where $A_0(t)\equiv \int_0^{2\pi R}\frac{dx}{2\pi R} A(t,\,x)$. For a derivation of eq.~(\ref{eq:eff_0p1lagr}) see e.g.~\cite{Draper:2018lyw}, that works with a $(4+1)$-dimensional theory. To obtain the effective potential derived from our $(1+1)$-dimensional theory one should multiply the effective potential in that paper by $(2\pi L/m)^{3/2}$ with $L=2\pi R$ to match our notation. Then, by solving the classical equations of motion derived from the Lagrangian~(\ref{eq:eff_0p1lagr}) to first order in the small quantity $e^{-2\pi mR}$ and identifying $E(t)=-\dot{A}_0$ we obtain
\begin{align}\label{eq:classical_osc}
E(t)=E+e\frac{m^{1/2}}{2\pi^2 R^{3/2}eE}e^{-2\pi Rm}\cos(2\pi R eEt)+{\cal O}(e^{-4\pi Rm})
\end{align}
which displays oscillations that, remarkably, have the same periodicity and whose amplitude has the same overall proportionality to $e^{-2\pi mR}$ as the result found in Section~\ref{subsec:rmgg1}. Actually, the oscillations in eq.~(\ref{eq:classical_osc}) match {\em exactly}, including ${\cal O}(1)$ factors, the result from the adiabatic component of $\langle \delta\dot{E}(t)\rangle_{(1)}$, presented in eq.~(\ref{eq:asymp_adiab}). This is not too surprising, as the effective potential in eq.~(\ref{eq:eff_0p1lagr}) has been obtained using the semiclassical approximation.  Going beyond the overall proportionality to $e^{-2\pi mR}$, however, the behavior found in the analysis of Section~\ref{subsec:rmgg1} above shows a different dependence on $m$ and $R$ than that of eq.~(\ref{eq:asymp_adiab}). This can be seen, for instance, from the fact the amplitude shown in the right panel of Figure~\ref{fig:mr12} is a decreasing function of $m$, whereas the result~(\ref{eq:asymp_adiab}) increases as $m^{1/2}$, once the overall proportionality to $e^{-2\pi mR}$ is factored out. This means that both the adiabatic contribution evaluated in Appendix~\ref{app:sigma} and the full one, obtained in Appendix~\ref{app:maincalc}, have the same overall sinusoidal behavior $\propto \cos(2\pi eE Rt)$, but a different amplitude. Inspection of those two terms in the $R={\cal O}(1)$, $m\gg 1$ regime shows that the they partially cancel, with the full component larger than the adiabatic one. It is possible that this difference is an artifact of the finiteness of the mass $m$ in our numerical calculations, and that the result will match exactly eq.~(\ref{eq:classical_osc}) in the limit $m\to\infty$.

Our analysis thus confirms, via a real time calculation performed in the full $(1+1)$-dimensional theory, the result presented in~\cite{Draper:2018lyw}, which in its turn explains, in terms of the compactified theory, the observation of~\cite{Brown:2015kgj} that, for small compactification radii, a new instanton with magnitude $\sim e^{-2\pi mR}$ would dominate the Schwinger effect. In particular, our analysis shows that the {\em net} rate of pair production is  proportional to $e^{-\pi \frac{m^2}{eE}}$ also for small compactification\footnote{It is worth stressing that our analysis has been performed to leading order in the in-in expansion, leading to a result $\langle \delta{E}(t)\rangle_{(1)}=e\,f_{(1)}(eE,\,m,\,R,\,t)$.  According to~\cite{Brown:2015kgj}, the new instanton should be associated to a final state with a lower electric field, but no matter particles. In our in-in language, this would correspond to a higher order term $\langle \delta{E}(t)\rangle_{(3)}=e^3\,f_{(3)}(eE,\,m,\,R,\,t)$ in the expansion~(\ref{inin_expansion}) that accounts for $\phi$ annihilation. It is thus not excluded that  a {\em net} contribution to the rate of pair production proportional to $ e^{-2\pi mR}$ might appear at that order in perturbation theory.
} radii, but for short timescales the effect associated to particle creation is subdominant with respect to the oscillations, with amplitude proportional to $e^{-2\pi mR}$ and frequency $R/eE$, induced by the virtual charged matter. 

\smallskip

To conclude, we have found real time formulae that allow to compute the change in the electric field due to  Schwinger effect in a compact space. The net rate of production of charged pairs, when evaluated on long times, is always proportional to $e^{-\pi\frac{m^2}{eE}}$, as in the non compact case. However, while in the limit of large compactification radii we recover the expected non compact result, in the regime of intermediate and small values of the dimensionless quantity $mR$ the quantity $\langle \delta{E}(t)\rangle_{(1)}$ shows a richer behavior. In particular, the steplike time dependence found in the regime of $mR\to 0$ breaks continuous time translations in a fashion that is analogous to the way bubble nucleation breaks Lorentz symmetry, and the oscillations found for $m^2\gg eE$, $R^2={\cal O}((eE)^{-1})$ can be explained as an effect of the virtual pairs of charged particles on the effective potential for the gauge field.

\acknowledgements We thank Patrick Draper, Gerald Dunne, David Kastor, Matt Kleban and Jennie Traschen for interesting discussions. This work is partially supported by the US-NSF grants PHY-1520292 and PHY-1820675.

\appendix

\section{Calculation for compact case}%
\label{app:maincalc}

In this Appendix we work out the steps that allow to go from the first term on the right hand side of eq.~(\ref{eq:implicit_sum}) to the right hand side of eq.~(\ref{eq:final_sum}) (except for the part proportional to the Bessel function $K_1$, that will be discussed in Appendix \ref{app:sigma} below).

We set $eE=1$, and focus on the sum
\begin{align}
{\cal S}=\frac{1}{R}\sum_{n=-\infty}^\infty \frac{e^{-\pi m^2/4}}{\sqrt{2}}\,\Big|D_{i\frac{m^2}{2}-\frac{1}{2}}\Big(-(n/R+t)\sqrt{2}e^{-i\pi/4}\Big)\Big|^2(n/R+t)\,,
\end{align}
so that 
\begin{eqnarray}
\langle \delta\dot{E}(t)\rangle_{(1)}=-\frac{e}{\pi}\,{\cal S}+\frac{e}{2\pi R}\,\sum_{n=-\infty}^{\infty} \frac{{n}/{R}+eEt}{\sqrt{({n}/{R}+eEt)^2+m^2}}\,.
\end{eqnarray}

We want to use the Mellin-Barnes representation, eq. 9.242.3 of~\cite{gradshteyn}, of the parabolic cylinder function
\begin{align}\label{eq:mellin}
D_{\nu}(z)=\frac{e^{-\frac{1}{4}z^{2}}z^{\nu}}{2\pi i\,\Gamma%
\left(-\nu\right)}\*\int_{-i\infty}^{i\infty}\Gamma\left(t\right)%
\Gamma\left(-\nu-2t\right)2^{t}z^{2t}{d}t,
\end{align}
that is valid for $|$Arg$(z)|<3\pi/4$. Given this restriction on Arg$(z)$, we have to treat the cases $(n/R+t)>0$ and $(n/R+t)<0$ separately, so that we write ${\cal S}\equiv {\cal S}_-+{\cal S}_+$ where $n$ in ${\cal S}_-$ goes from $-N_-$ to $[-Rt]-1$ (we define here the integer part  in such a way that $[-x]=-[x]$), and in ${\cal S}_+$ goes from $[-Rt]$ to $N_+$. Here, $N_\pm>0$ are regulators that we will eventually send to infinity.

Let us first consider ${\cal S}_-$, where $-(n/R+t)\sqrt{2}\equiv |z|>0$. In this case we can use eq.~(\ref{eq:mellin}) right away. It is convenient to write the integral~(\ref{eq:mellin}) as an asymptotic series on the poles at $t=-j$, $j=0,\,1,\,2,\,...$
\begin{align}\label{eq:mellin_series}
&D_{ia-1/2}(|z|e^{-i\pi/4})=\frac{e^{\frac{i}{4}|z|^{2}}\,|z|^{ia-\frac{1}{2}}\,e^{\pi a/4+i\pi/8}}{\Gamma%
\left(\frac{1}{2}-ia\right)}\sum_{j=0}^\infty%
\frac{(-1)^j}{j!}
\Gamma\left(\tfrac{1}{2}-ia+2j\right)2^{-j}\,|z|^{-2j}e^{i\pi j/2}\,,
\end{align}
that allows us to write, after relabelling $n\to -n$,
\begin{align}
&{\cal S}_-= -\frac{\cosh(\pi m^2/2)}{2\pi R}\sum_{n'=[Rt]+1}^{N_-}\sum_{j,k=0}^\infty
\frac{(-1)^j}{j!}\frac{(-1)^k}{k!}
\Gamma\left(\tfrac{1}{2}-i\tfrac{m^2}{2}+2j\right)\Gamma\left(\tfrac{1}{2}+i\tfrac{m^2}{2}+2k\right)\left(\frac{R}{2}\right)^{2j+2k}\frac{e^{i\pi (j-k)/2}}{(n'-Rt)^{2j+2k}}\,.
\end{align}
Then, we separate the sum into the components with $j=k=0$, $j=0$ and $k\ge 1$, $j\ge 1$ and $k=0$, and $j,k\ge 1$:
\begin{align}
&{\cal S}_-= -\left\{\frac{1}{2R}\sum_{n'=[Rt]+1}^{N_-}1\right\}-2\Re\left\{\frac{1}{2 R}\sum_{n'=[Rt]+1}^{N_-}\sum_{k=1}^\infty
\frac{(-1)^k}{k!}
\frac{\Gamma\left(\tfrac{1}{2}+i\tfrac{m^2}{2}+2k\right)}{\Gamma\left(\tfrac{1}{2}+i\tfrac{m^2}{2}\right)}\left(\frac{R}{2}\right)^{2k}\frac{1}{(n'-Rt)^{2k}}e^{i\pi (-k)/2}\right\}\nonumber\\
&-\frac{1}{2 R}\sum_{n'=[Rt]+1}^{N_-}\sum_{j,k=1}^\infty
\frac{(-1)^j}{j!}\frac{(-1)^k}{k!}
\frac{\Gamma\left(\tfrac{1}{2}+i\tfrac{m^2}{2}+2k\right)}{\Gamma\left(\tfrac{1}{2}+i\tfrac{m^2}{2}\right)}\frac{\Gamma\left(\tfrac{1}{2}-i\tfrac{m^2}{2}+2j\right)}{\Gamma\left(\tfrac{1}{2}+i\tfrac{m^2}{2}\right)}\left(\frac{R}{2}\right)^{2j+2k}\frac{e^{i\pi (j-k)/2}}{(n'-Rt)^{2j+2k}}\,.
\end{align}
This allows us to isolate the divergence in the limit $N_-\to\infty$, that appears only in the first term of the equation above.  We can now send $N_-\to\infty$ in the remaining terms, shift the summation variable $n$ by  $[Rt]+1$, and write
\begin{align}\label{eq:final_sminus}
&{\cal S}_-=-\frac{N_--[Rt]-1}{2R}- \frac{1}{2R}\sum_{n=0}^{\infty}\left[\left|{\cal G}+1\right|^2-1\right]\,,
\end{align}
where
\begin{align}
&{\cal G}=\sum_{j=1}^\infty
\frac{(-1)^j}{j!}
\frac{\Gamma\left(\tfrac{1}{2}-i\tfrac{m^2}{2}+2j\right)}{\Gamma\left(\tfrac{1}{2}-i\tfrac{m^2}{2}\right)}\left(\frac{1}{2}\right)^{j}\frac{1}{[\sqrt{2}(n+[Rt]+1-Rt)/R]^{2j}}e^{i\pi j/2}\,,
\end{align}
that, using again eq.~(\ref{eq:mellin_series}), gives
\begin{align}\label{eq:calG}
\left|{\cal G}+1\right|^2=\sqrt{2}\,\frac{n+[Rt]+1-Rt}{R}\,e^{-\pi m^2/4}\,\left|D_{im^2/2-1/2}\left(\sqrt{2}\,\frac{n+[Rt]+1-Rt}{R}e^{-i\pi/4}\right)\right|^2\,.
\end{align}

Next, we have to take care of ${\cal S}_+$, where the phase of the argument of the parabolic cylinder function  is precisely $3\pi/4$, so that the expression~(\ref{eq:mellin}) is not directly applicable. In order to use eq.~(\ref{eq:mellin}) we have to first apply the third of eqs.~9.248.1 of~\cite{gradshteyn}:
\begin{align}
D_p(z)=e^{i\pi p}D_p(-z)+\frac{\sqrt{2\pi}}{\Gamma(-p)}e^{i(p+1)\pi/2}D_{-p-1}(-iz)\,,
\end{align}
so that ${\cal S}_+$ reads
\begin{align}
{\cal S}_+&=\frac{1}{2\,R}\sum_{n=[-Rt]}^{N_+}\Bigg[e^{-5\pi m^2/4}\,\Big|D_{i\frac{m^2}{2}-\frac{1}{2}}(|n/R+t|\sqrt{2}e^{-i\pi/4})\Big|^2\sqrt{2}(n/R+t)\nonumber\\
&+e^{-3\pi m^2/4}\,\Big|\frac{\sqrt{2\pi}}{\Gamma(-i\frac{m^2}{2}+\frac{1}{2})}D_{-i\frac{m^2}{2}-\frac{1}{2}}(|n/R+t|\sqrt{2}e^{i\pi/4})\Big|^2\sqrt{2}(n/R+t)\nonumber\\
&+2\,\sqrt{2}\,e^{-\pi m^2}\Re\left\{-iD_{i\frac{m^2}{2}-\frac{1}{2}}(|n/R+t|\sqrt{2}\,e^{-i\pi/4})\frac{\sqrt{2\pi}\,e^{-i\pi/4}}{\Gamma(i\frac{m^2}{2}+\frac{1}{2})}D_{i\frac{m^2}{2}-\frac{1}{2}}(|n/R+t|\sqrt{2}e^{-i\pi/4})(n/R+t)\right\}\Bigg]\,.
\end{align}
The summand in the third line of this expression is quickly oscillating, and we neglect it. The terms in the first two lines can be treated in a way that is analogous to the one that led to eqs.~(\ref{eq:final_sminus}) and~(\ref{eq:calG}).

We thus obtain the desired result
\begin{align}
&{\cal S}=\frac{1}{2R}\Big\{(1+2\,e^{-\pi m^2})\sum_{n=0}^\infty \left[\frac{\sqrt{2}(n+Rt-[Rt])}{R}\,e^{-\pi m^2/4}\left|D_{-1/2+im^2/2}\left(\frac{\sqrt{2}(n+Rt-[Rt])}{R}\,e^{-i\pi/4}\right)\right|^2-1\right]\nonumber\\
&-\sum_{n=0}^\infty \left[\frac{\sqrt{2}(n+1-Rt+[Rt])}{R}\,e^{-\pi m^2/4}\left|D_{-1/2+im^2/2}\left(\frac{\sqrt{2}(n+1-Rt+[Rt])}{R}\,e^{-i\pi/4}\right)\right|^2-1\right]\nonumber\\
&+(2+2\,e^{-\pi m^2})\,[Rt]+(1+2\,e^{-\pi m^2})\left(1+N_+\right)-N_-\Big\}\,,
\end{align}
where the sums over $n$ are finite, and the divergences in the terms containing $N_\pm$ have been isolated.

\section{The vacuum contribution}%
\label{app:sigma}

The contribution to $\langle \delta\dot{E}(x,\,t)\rangle_{(1)}$ from the vacuum can be read from eq.~(\ref{eq:implicit_sum}) and takes the form
\begin{align}
\frac{e}{2\pi R}\sum_{n=-N_-}^{N_+}\frac{\frac{n}{R}+eEt}{\sqrt{(\frac{n}{R}+eEt)^2+m^2}}\,.
\end{align}
This sum is divergent as $N_\pm\to+\infty$, but its derivative with respect to $Rt$ is convergent, so that we can take the limit $N_\pm\to+\infty$ after differentiation and obtain (after setting $eE=1$)
\begin{align}
&\frac{e}{2\pi R}\frac{\partial}{\partial(Rt)}\left\{\sum_{n=-N_-}^{N_+} \frac{n+Rt}{\sqrt{(n+Rt)^2+(Rm){}^2}}\right\}=\frac{e}{2\pi R}\sum_{n=-\infty}^{\infty} \frac{(mR){}^2}{[(n+Rt)^2+(mR){}^2]^{3/2}}\nonumber\\
&=\frac{e}{2\pi R}\frac{(mR){}^2}{\Gamma(3/2)}\int_0^\infty dw\,w^{1/2}\sum_{n=-\infty}^{\infty} e^{-w[(n+Rt)^2+(mR){}^2]}=\frac{e}{\pi R}\,(mR){}^2\int_0^\infty dw\,e^{-w\,(mR){}^2}\theta_3(\pi Rt,\,e^{-\pi^2/w})\,,
\end{align}
where $\theta_3$ denotes the third Jacobi $\theta$ function. Using the representation $\theta_3(u,\,q)=1+2\sum_{n=1}^\infty q^{n^2}\,\cos(2nu)$, we can write
\begin{align}
&\frac{e}{2\pi R}\frac{\partial}{\partial(Rt)}\left\{\sum_{n=-N_-}^{N_+} \frac{n+Rt}{\sqrt{(n+Rt)^2+(Rm){}^2}}\right\}=\frac{e}{2\pi R}\left[2+4(mR){}^2\sum_{n=1}^\infty \cos(2\pi n Rt)\,\frac{2\pi n}{mR}\,K_1(2\pi n mR)\right]\,,
\end{align}
that is a more transparent sum. Integrating back in $d(Rt)$, we obtain
\begin{align}
\frac{e}{2\pi R}\sum_{n=-N_-}^{N_+} \frac{n+Rt}{\sqrt{(n+Rt)^2+(Rm){}^2}}=\frac{e}{\pi R}\left[Rt+2mR{}\sum_{n=1}^\infty \sin(2\pi n tR)\,K_1(2\pi n mR)+{\rm {constant}}\right]\,,
\end{align}
where the constant is generally divergent, as it depends on the cutoffs $N_\pm$.

In the limit $mR\to 0$ we can use the small argument approximation of the Bessel function, $K_1(x)\simeq 1/x$, to write
\begin{align}
\frac{e}{2\pi R}\sum_{n=-N_-}^{N_+} \frac{n+Rt}{\sqrt{(n+Rt)^2+(Rm){}^2}}\Bigg|_{mR\ll 1}&\simeq \frac{e}{\pi R}\left[ Rt+\sum_{n=1}^\infty \frac{\sin(2 \pi n tR)}{n\,\pi}+{\rm {constant}}\right]\nonumber\\
&= \frac{e}{\pi R}\left[Rt+\frac{1-2tR}{2}+{\rm {constant}}\right]={\rm {constant}}\,,
\end{align}
where the approximation is valid for $tR\gg mR$. Of course, for $tR=0$ and $tR=1$, $\sin(2\pi ntR)=0$, so 
\begin{align}
\frac{e}{2\pi R}\sum_{n=-N_-}^{N_+} \frac{n+Rt}{\sqrt{(n+Rt)^2+(Rm){}^2}}\Bigg|_{tR=1}-\frac{e}{2\pi R}\sum_{n=-N_-}^{N_+} \frac{n+Rt}{\sqrt{(n+Rt)^2+(Rm){}^2}}\Bigg|_{tR=0}=\frac{e}{\pi R}\,,
\end{align}
which implies
\begin{align}\label{eq:deltaSsmallR}
\frac{e}{2\pi R}\sum_{n=-N_-}^{N_+} \frac{n+Rt}{\sqrt{(n+Rt)^2+(Rm){}^2}}\Bigg|_{mR\ll 1}\simeq \frac{e}{\pi R}[tR]+{\rm {constant}}\,.
\end{align}

On the other hand, in the large $2\pi mR$ limit, we use the asymptotics $K_1(x)\simeq \sqrt{\frac{\pi}{2x}}e^{-x}$ to keep only the first term in the series, obtaining
\begin{align}\label{eq:asymp_adiab}
\frac{e}{2\pi R}\sum_{n=-N_-}^{N_+} \frac{n+Rt}{\sqrt{(n+Rt)^2+(Rm){}^2}}\Bigg|_{mR\gg 1}=\frac{e}{\pi R}\left[Rt+\sqrt{{m}{R}}\,\sin(2\pi tR)\,e^{-2\pi mR}+{\rm {constant}}\right]\,.
\end{align}

\end{document}